# Elastic wave dispersion in layered media with suture joints: influence of structural hierarchy and viscoelasticity

Federica Ongaro[1,2], Federico Bosia[3] and Nicola M. Pugno[1,4]

[1]Department of Civil, Environmental and Mechanical Engineering, Laboratory for Bio-inspired, Bionic, Nano, Meta Materials and Mechanics, University of Trento, Italy

[2]Department of Weapon Systems and Ballistics, Royal Military Academy, 30 Avenue de la Renaissance, 1000 Brussels, Belgium [3]Department of Applied Science and Technology, Politecnico di Torino, Torino, TO, Italy

[4]School of Engineering and Materials Science, Queen Mary University of London, UK

**Author for correspondence:**

Nicola M. Pugno

e-mail: nicola.pugno@unitn.it

**Abstract:**

Suture joints contribute to the exceptional combination of stiffness, strength, toughness and efficient load bearing and transmission of many biological structures like the cranium or ammonite fossil shells. However, their role in the attenuation of vibrations and effect on dynamic loads is less clear. Moreover, the self-similar hierarchical geometry often associated with suture joints renders its treatment with standard numerical approaches computationally prohibitive. To address this problem, this paper investigates the dynamic response of periodic layered media with suture joints using an analytical approach based on material homogenization. A general trapezoidal suture geometry is considered together with the fundamental ingredients of hierarchy and viscoelasticity. The Spectral Element Method and Bloch theorem are used to derive the dispersion relation and band diagram of the system, including propagating and evanescent dispersion modes. A strong influence of the suture morphology and material properties emerges, and the analysis reveals an important advantage of adding hierarchy, i.e. the possibility of simultaneously obtaining wider bandgaps and their shift to higher frequencies. A synergy between hierarchy and structure is also observed, providing superior levels of wave attenuation. These findings suggest a possible design concept for bioinspired devices with efficient and tailorable wave attenuation properties.

# 1. Introduction

The role of suture joints in the superior mechanical properties of biological systems like the cranium [1], ammonite fossil shells [2] and woodpecker beak [3], among others, is currently a hot research topic. From a mechanical point of view, suture joints are composite structures typically including two interdigitating stiff components (the teeth) joined by a thin, more compliant interface layer. This particular configuration endows the biological system not only with a high level of flexibility to accommodate vital functions, such as growth, respiration and protection from predation [4–7], but also a more efficient way to bear and transmit loads, absorb energy and generally obtain an exceptional combination of stiffness, strength and toughness [8–12].

Starting from the pioneering contributions in [5,12,13], dealing with the geometric characterization of different types of biological suture joints, many authors extensively studied the mechanics of such wavy structures and a vast literature has flourished in recent years. Noteworthy contributions are reported in [14,15], where the functional significance of the high sinuosity and complexity of ammonoid suture patterns in decreasing stresses and deformations is explained, and in [7,16], where the role of geometry on the suture mechanical behaviour and effective properties, including stiffness, strength, fracture toughness and failure, is discussed. The analysis is then extended in [11,17] for the wavy-patterned suture joints of the common millet (*Panicum miliaceum*) seedcoat. The fundamental role played by the sutures in resisting indentation loads is demonstrated by means of an integrated experimental, numerical and analytical investigation [17] and the influence of the suture geometric and mechanical parameters on the effective properties is clarified [11]. Wavy-patterned sutures are also investigated in [18], where an application of a multiscale fracture analysis illustrates how the effective fracture resistance and damage tolerance are affected by the suture waveform. Based on finite-element analysis, a better understanding of the mechanical behaviour of suture joints is presented in [19–21], with an investigation on how the suture morphology, i.e. interdigitating factor and connectivity, impacts the total strain energy absorption capability of bone-suture structures. A number of experimental and numerical investigations are proposed in [5,13,22,24], where it emerges that cranial sutures enhance energy absorption during impact loading [5] and increase the resistance against interlaminar delamination and crack propagation [24].

In line with [7,16], analytical relations between the effective properties and the geometric and mechanical features of hierarchical suture joints with a triangular profile are proposed in [2], where an increase in the stiffness-to-density ratio is obtained by increasing the number of hierarchical levels. The importance of the hierarchical-like morphology is also pointed out in [4] for the case of cranial suture joints. Finally, the damping performances of suture structures with a viscoelastic interface layer are considered in [8], where the influence of the suture configuration on the damping properties like loss factor, loss modulus and storage modulus is explored.

From the suggested examples, which encompass the majority of current work in the field, it can be said that, in the literature, many studies deal with the mechanical characterization of biologically inspired suture joints, with particular attention on the role of the suture geometry on the overall behaviour. Surprisingly, very few investigations on suture joints where one or both phases display a viscoelastic behaviour, as commonly happens in biology, are currently available. In addition, the scope of the above examples is limited to the mechanical response of sutures under static or, in very few cases, cycling loading conditions. In reality, dynamic loading situations often occur, as in the emblematic case of the cranial suture structure, which protects various fragile organs like the brain [25], in particular in cases like the woodpecker skull, where a high vibration damping ability is required [3]. Thus, it would be interesting to discuss how the morphological and mechanical features of suture joints influence their dynamic response. Some efforts in this direction have been proposed in [25–28], where a finite-element model has been established to investigate how cranial sutures respond to impulsive loading in terms of crack behaviour [29], energy absorption and stress wave mitigation [25–27]. However, as far as we know, an analytical study on the dynamic response of composite materials with suture joints is still lacking and, in particular, the three fundamental aspects of dynamics, hierarchy and viscoelasticity remain to be considered simultaneously in the context of suture joints. To provide a contribution in this incomplete field of research, this paper deals with the analysis of wave propagation in two-dimensional periodic layered media with suture joints, in a configuration inspired by the previously mentioned common millet seedcoat. The general case of a trapezoidal suture profile is considered, allowing us to explore how the suture geometry and material properties, i.e. elastic or viscoelastic behaviour, affect the dynamic response in terms of wave attenuation performance. The effects of adding some levels of hierarchy are also illustrated. From a numerical point of view, this addition leads to a computationally highly expensive problem so that an analytical approach based on homogenization is adopted. The work is organized in five sections, including this Introduction. Initially, the problem under consideration is defined in §2, where the extension of the theory to hierarchical sutures is also presented. The dynamic problem is treated in §3 by adopting the spectral element method in conjunction with the Bloch theorem. This provides the dispersion relation of the investigated system, from which the band diagram is obtained. Two different scenarios are examined and discussed in §4: a non-dissipative linear elastic response of the medium and a dissipative viscoelastic one. In both cases, some considerations about the influence of the suture morphology, i.e. geometric profile, hierarchical or non-hierarchical configuration, are presented. Section 6 summarizes the main findings.

## 2. Layered media with suture joints

### (a) Problem setting

Let us focus on the two-dimensional layered medium illustrated in figure 1, obtained by periodically alternating two different layers $\pounds_1$ and $\pounds_2$ having thickness, respectively, of $d_1$ and $d_2$. $\pounds_1$ is assumed to be a homogeneous, linear elastic layer with mass density $\rho_1$, shear modulus $G_1$ and Young's modulus $E_1$. Conversely, $\pounds_2$ is a suture layer, i.e. a composite layer including two trapezoidal interdigitating stiff phases: the teeth, joined by a thin compliant element, the interface layer, along the seam line (figures 1 and 2). Both the teeth and the interface layer are homogeneous and perfectly bonded at the slant interfaces. Regarding their mechanical response, two different situations are considered: (i) linear elastic behaviour for both phases, (ii) linear elastic teeth and viscoelastic interface layer. In the latter, in particular, the Kelvin–Voigt model has been used, described by [8]:

$$\begin{cases} \tau_L = G_L \gamma + \eta \dot{\gamma} & \text{shear behaviour} \\ \sigma_L = E_L \varepsilon + \xi \dot{\varepsilon} & \text{normal behaviour} \end{cases} \quad (2.1)$$

with $G_L$ the layer shear modulus, $E_L$ Young's modulus, $\eta$ the viscosity coefficient for shear deformation $\gamma$, $\xi$ the viscosity coefficient for normal deformation $\varepsilon$ and $(\dot{\cdot}) := d(\cdot)/dt$.

To simplify the analysis, $\pounds_2$ is approximated with an equivalent continuum (figure 1, right) assumption that, in the dynamic context presented in §3, restricts our approach to the case of long wavelengths. Specifically, by focusing on the interdigitating area in figure 2 and by applying, in the viscoelastic case, the elastic–viscoelastic correspondence principle [7,30], it emerges that the effective properties of $\pounds_2$ are given by (see appendix A for further details):

$$E_2 = \left[ \frac{\lambda_S - 2h_L}{\lambda_S E_T} + \frac{2h_L}{\lambda_S} \left( \frac{\cos^2 \alpha_S \sin^2 \alpha_S}{\tilde{G}_L} + \frac{\cos^4 \alpha_S}{\tilde{E}_L} \right) \right]^{-1}, \quad (2.2)$$

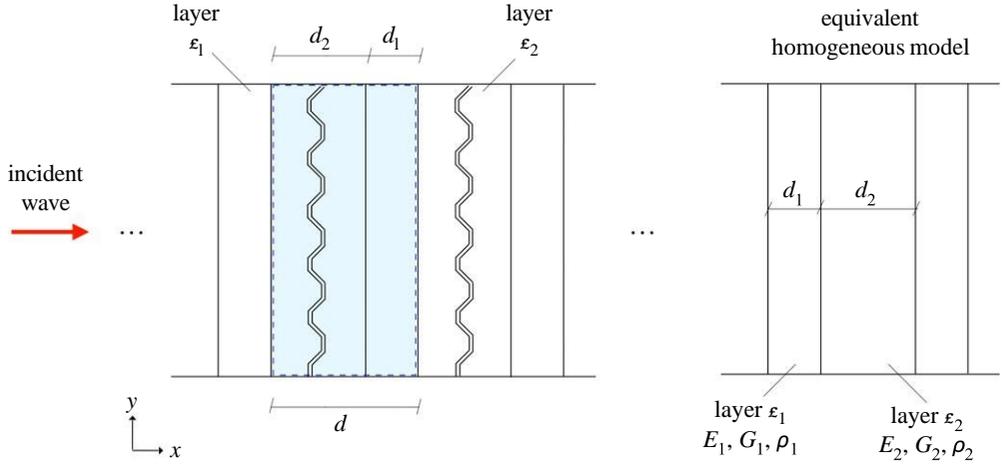

**Figure 1.** Layered media with suture joints: configuration considered and corresponding equivalent homogeneous model.

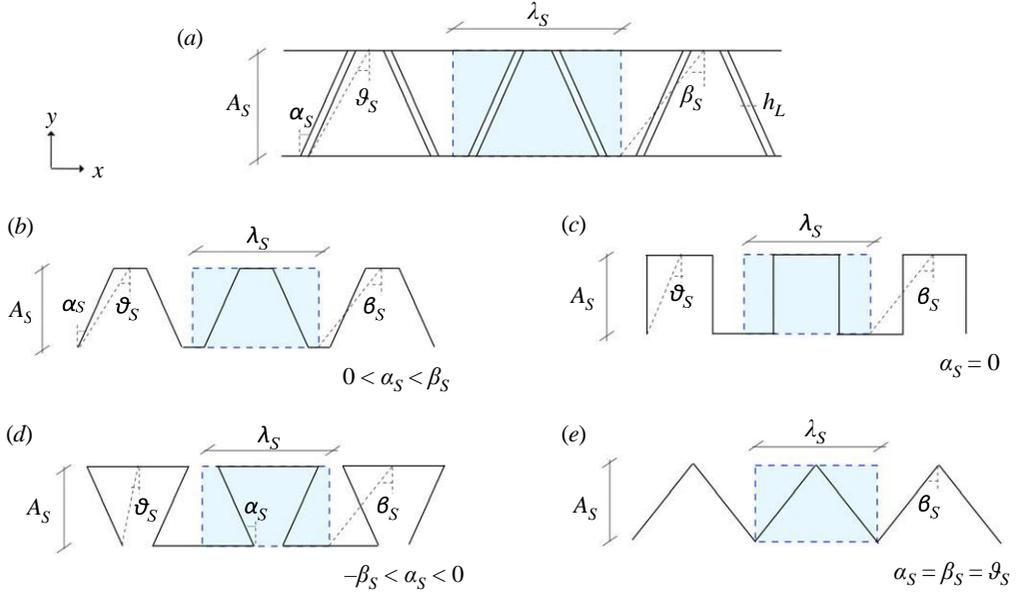

**Figure 2.** Schematic of the trapezoidal suture area in layer $\pounds_2$, indicating (*a*) the most relevant geometric parameters and the geometric profiles corresponding to different values of the angles $\alpha_S$, $\beta_S$, $\vartheta_S$: (*b*) trapezoidal, (*c*) rectangular, (*d*) anti-trapezoidal and (*e*) triangular.

and

$$G_2 = \frac{(\lambda_S - 2h_L)\Phi_1}{((\Phi_1/4\Phi_2)(3 + (5\Phi_2/4E_T \tan^2 \beta_S)) + (\lambda_S \tan \alpha_S/2A_S)((2h_L/\lambda_S \psi_{\alpha\beta}) + (3\tilde{G}_L/2\Phi_2)(1 + (3\Phi_2/4E_T \tan^2 \beta_S))))\lambda_S}, \quad (2.3)$$

with $E_2$ the layer Young's modulus in the longitudinal direction and $G_2$ the shear modulus. In equations (2.2) and (2.3), the parameters $A_S$ and $\lambda_S$ are, respectively, the amplitude and wavelength of the suture profile, $h_L$ the thickness of the interface layer, $\alpha_S$ and $\beta_S$ the angles defining the suture geometry, i.e. trapezoidal, rectangular, triangular and anti-trapezoidal (figure 2), $E_T$ and $G_T$ Young's modulus and shear modulus of the teeth, $\tilde{E}_L := E_L + i\omega\xi$, $\tilde{G}_L :=$

$G_L + i\omega\eta$, and

$$\psi_{\alpha\beta} := \frac{\tan \alpha_S}{\tan \beta_S}, \quad \Phi_1 := \tilde{G}_L \tan^2 \alpha_S + \tilde{E}_L, \quad \Phi_2 := \frac{G_T(\lambda_S - 2h_L) + 2\tilde{G}_L h_L}{\lambda_S}. \quad (2.4)$$

Finally, the rule of mixtures provides the effective mass density of $\pounds_2$:

$$\rho_2 = (1 - \varphi_T)\rho_L + \varphi_T \rho_T, \quad (2.5)$$

with $\rho_T$ and $\rho_L$, in turn, the mass density of the teeth and of the interface layer, and

$$\varphi_T = 1 - \frac{2h_L}{\lambda_S} \quad (2.6)$$

the volume fraction of the teeth [7].

## (b) Hierarchical extension

Let us imagine modifying the layered medium in figure 1 by replacing the homogeneous interface layer in $\pounds_2$ with a structural element having the same composite configuration of $\pounds_2$, i.e. two stiff components articulated via a thin suture layer. If we iterate this modification at successively smaller length scales, we obtain the hierarchical arrangement shown in figure 3b, representing the general case of a level-$[n]$ hierarchical suture element, where $n$, the hierarchical order, is defined as the numbers of levels of scale displaying a recognized structure [31]. In terms of the overall layered system of figure 3a, $\pounds_1$ is assumed to be an isotropic, linear elastic layer as in §2a, with mass density $\rho_1$, Young's modulus $E_1$ and shear modulus $G_1$. Again, $\pounds_2$ is approximated with an equivalent continuum and its effective properties in the general case of $n$ levels of hierarchy, Young's modulus in the layer longitudinal direction, $E_2^{[n]}$, and shear modulus, $G_2^{[n]}$, take the form:

$$E_2^{[n]} = \left[ \frac{\lambda_S^{[n]} - 2h_L^{[n]}}{\lambda_S^{[n]} E_T^{[n]}} + \frac{2h_L^{[n]}}{\lambda_S^{[n]}} \left( \frac{\cos^2 \alpha_S^{[n]} \sin^2 \alpha_S^{[n]}}{\tilde{G}_L^{[n]}} + \frac{\cos^4 \alpha_S^{[n]}}{\tilde{E}_L^{[n]}} \right) \right]^{-1} \quad (2.7)$$

and

$$G_2^{[n]} = ((\frac{}{\Phi_1^{[n]}/4\Phi_2^{[n]}})(\frac{}{(3 + 5\Phi_1^{[n]}/4E_L^{[n]} \tan^2 \beta_S^{[n]}})) \frac{(\lambda_S^{[n]} - 2h_L^{[n]}) \Phi_1^{[n]}/\lambda_S^{[n]}}{}$$
$$+ \lambda_S^{[n]} \tan \alpha_S^{[n]}/2A^{[n]})((2h_L^{[n]}/\lambda_S^{[n]}\psi_{\alpha\beta}^{[n]} + (3\tilde{G}_L^{[n]}/2\Phi_2^{[n]})(1 + 3\Phi_2^{[n]}/4E_L^{[n]} \tan^2 \beta_S^{[n]}))))$$
$$\quad (2.8)$$

The two sets of parameters $A^{[n]}, \lambda_S^{[n]}, \alpha_S^{[n]}, \beta_S^{[n]}, h_S^{[n]}$ and $E^{[n]}, G^{[n]}, \tilde{E}_L^{[n]}, \tilde{G}_L^{[n]}$, respectively, specify the geometric and mechanical characteristics of the level-$[n]$ suture (figure 3b), $\psi_{\alpha\beta}^{[n]}, \Phi_1^{[n]}$ and $\Phi_2^{[n]}$ given by

$$\psi_{\alpha\beta}^{[n]} := \frac{\tan \alpha_S^{[n]}}{\tan \beta_S^{[n]}}, \quad \Phi_1^{[n]} := \tilde{G}_L^{[n]} \tan^2 \alpha_S^{[n]} + \tilde{E}_L^{[n]} \quad (2.9)$$

and

$$\Phi_2^{[n]} := \frac{G_T^{[n]}(\lambda_S^{[n]} - 2h_L^{[n]}) + 2\tilde{G}_L^{[n]} h_L^{[n]}}{\lambda_S^{[n]}}.$$

The effective mass density of $\pounds_2$, $\rho_2^{[n]}$, is provided by the rule of mixtures:

$$\rho_2^{[n]} = (1 - \varphi_T^{[n]})\rho_L^{[n]} + \varphi_T^{[n]} \rho_T^{[n]}, \quad (2.10)$$

with $\varphi_T^{[n]}$ the volume fraction of the teeth and $\rho_T^{[n]}, \rho_L^{[n]}$, respectively, the mass density of the teeth and of the interface layer in the level-$[n]$ system.

It should be noted that the above expressions are derived by applying the same methodology of §2a in conjunction with the assumption that, at each hierarchical level, the size of the interface layer microstructure is fine enough to be negligible with respect to the underlying large architecture. This allows us to treat the interface layer as a continuum with effective properties having the same form of equations (2.7), (2.8) and (2.10). To elucidate the procedure, let us focus on the level-[*n*] suture joint in figure 3*b*. As can be seen, the interface layer is a level-[*n* − 1] element so that its mechanical properties, denoted by $\tilde{E}_L^{[n]}$ and $\tilde{G}_L^{[n]}$, and mass density, $\rho^{[n]}$, coincide with the effective moduli and the effective mass density, respectively, of the [*n* − 1]th hierarchical suture, derived from equations (2.7), (2.8) and (2.10) by substituting the parameters $(\cdot)^{[n]}$ with the parameters $(\cdot)^{[n-1]}$.

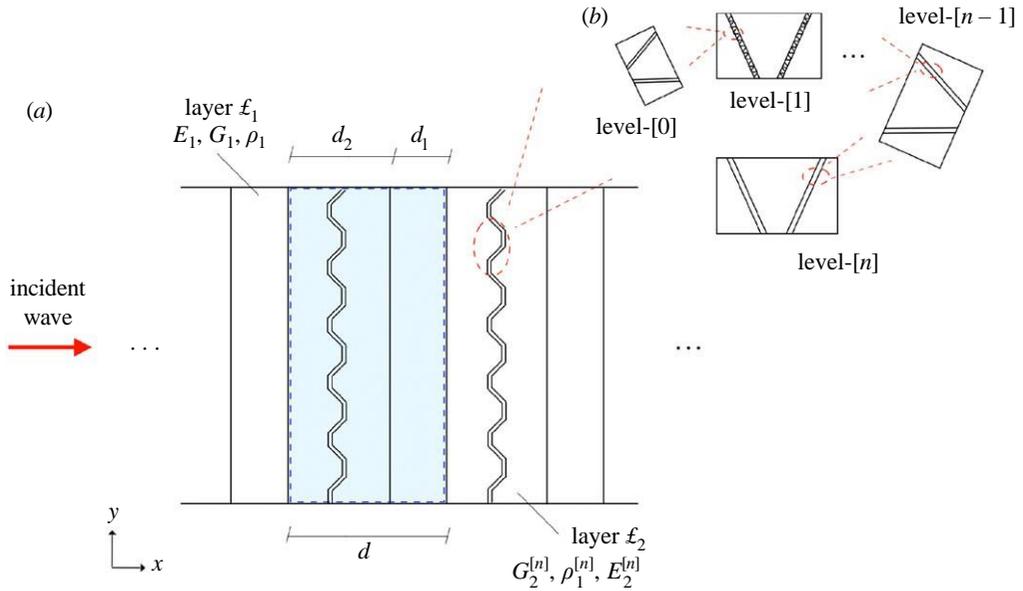

**Figure 3.** Hierarchical-layered media with suture joints: (*a*) examined system and (*b*) focus on the hierarchical arrangement of the suture in layer $£_2$.

## 3. Wave propagation in layered media with suture joints: basic concepts and assumptions

Consider now a unidirectional time-harmonic plane wave propagating along the medium in figures 1 and 3 normally to the layer interfaces (*P*-wave).

Two different methods can be used to investigate wave propagation along a layered structure. The first method is the *Direct Method* or *Transfer Matrix Method* that consists in initially establishing, for each layer, the so-called transfer matrix that relates the wavefields on the two boundaries of the layer and then, by multiplication, obtaining the transfer matrix of the full chain of periodic layers that, inserted in a classical eigenvalue problem, gives the dispersion relation [32]. The second technique is the *Spectral Element Method* [32] where, contrary to the previous one, the dispersion relation is obtained by considering a single spectral element or unit cell, formed by coupling two successive layers. The continuity of stresses and displacements at the interface is then imposed, together with the Bloch–Floquet periodicity condition [33] at the boundaries of the unit cell. The latter assumption, in particular, is typical for time-harmonic waves travelling within periodic media. Given the periodicity and infinite extent of the considered medium, in the present paper, in line with [34,35], the *Spectral Element Method* has been used.

In our case, the unit cell coincides with the combined bilayer system $£_1 + £_2$ having thickness $d := d_1 + d_2$ (figures 1 and 3) and subjected, by assumption, to time-harmonic vibrations. For

infinitesimal deformations, the governing equation of motion reads (compression waves are considered):

$$E_j \frac{\partial^2 u_j(x,t)}{\partial x^2} = \rho_j \frac{\partial^2 u_j(x,t)}{\partial t^2}, \quad j=1,2 \qquad (3.1)$$

where $E_j$, $\rho_j$ and $u_j(x,t)$ are, respectively, Young's modulus, mass density and horizontal displacement of the layer $\pounds_j$, $j = 1, 2$.

For the sake of clarity, in the case of a hierarchical arrangement of the layer $\pounds_2$ (cf. §2b), equation (3.1) takes the form

$$\begin{cases} E_1 \dfrac{\partial^2 u_1(x,t)}{\partial x^2} = \rho_1 \dfrac{\partial^2 u_1(x,t)}{\partial t^2}, & \text{layer } \pounds_1 \\ E_2^{[n]} \dfrac{\partial^2 u_2^{[n]}(x,t)}{\partial x^2} = \rho_2^{[n]} \dfrac{\partial^2 u_2^{[n]}(x,t)}{\partial t^2}, & \text{layer } \pounds_2 \end{cases} \qquad (3.2)$$

with $E_1$, $\rho_1$, $u_1(x,t)$ and $E_2^{[n]}$, $\rho_2^{[n]}$, $u_2^{[n]}$, in turn, Young's modulus, mass density and displacement of layers $\pounds_1$ and $\pounds_2$ in the general case of $n$ orders of hierarchy.

In the following, to facilitate reading, the more simple and compact notation of equation (3.1) will be used. However, the adopted methodology and the derived considerations apply for both the non-hierarchical and hierarchical system, even if the apex $[n]$ is not specified.

Let us thus focus on the second-order differential equation (3.1) whose solution, providing the displacement field within each layer, can be expressed as

$$u_j(x,t) = A_j e^{i(k_j x - \omega t)} + B_j e^{i(-k_j x - \omega t)}, \quad j=1,2, \qquad (3.3)$$

where $i := \sqrt{-1}$ is the imaginary unit, $A_j$ and $B_j$ the wave amplitudes (forward and backward travelling waves are considered), $\omega$ the circular frequency of the time-harmonic plane wave propagating along the medium, $k_j$ the layer specific wavenumber, given by

$$k_j = \frac{\omega}{\sqrt{E_j/\rho_j}}, \quad j=1,2. \qquad (3.4)$$

The corresponding stress field takes the form

$$\begin{aligned} \sigma_j(x,t) &= E_j \frac{\partial u_j(x,t)}{\partial x} \\ &= E_j(ik_j(A_j e^{i(k_j x - \omega t)}) - ik_j(B_j e^{i(-k_j x - \omega t)})), \quad j=1,2. \end{aligned} \qquad (3.5)$$

Assuming that the two layers are perfectly bonded to each other, the following conditions enforcing the continuity of displacements and stresses, i.e. the primary field variable and its spatial derivative, can be imposed at the layers interface $x = \tilde{x}$ [34,36]:

$$\begin{cases} f.u(x,t)|_{x=\tilde{x}} = 0 \\ f.\sigma(x,t)|_{x=\tilde{x}} = 0 \end{cases} \rightarrow \begin{cases} u_1(x=\tilde{x},t) = u_2(x=\tilde{x},t) \\ \sigma_1(x=\tilde{x},t) = \sigma_2(x=\tilde{x},t) \end{cases} \qquad (3.6)$$

with $f.f(x)|_{x=\tilde{x}} := \lim_{\delta \to 0}[f(\tilde{x}+\delta) - f(\tilde{x}-\delta)]$ the jump of the function $f$ at $x = \tilde{x}$. Also, assuming that the stresses and displacements satisfy the Bloch periodicity condition at the unit cell outer boundaries (figure 4), leads to

$$\begin{cases} u(x+d,t) = u(x,t) e^{i(qd-\omega t)} \\ \sigma(x+d,t) = \sigma(x,t) e^{i(qd-\omega t)} \end{cases} \rightarrow \begin{cases} u_2(x=d_2,t) = u_1(x=-d_1,t) e^{i(qd-\omega t)} \\ \sigma_2(x=d_2,t) = \sigma_1(x=-d_1,t) e^{i(qd-\omega t)} \end{cases} \qquad (3.7)$$

with $q$ the Bloch wavevector.

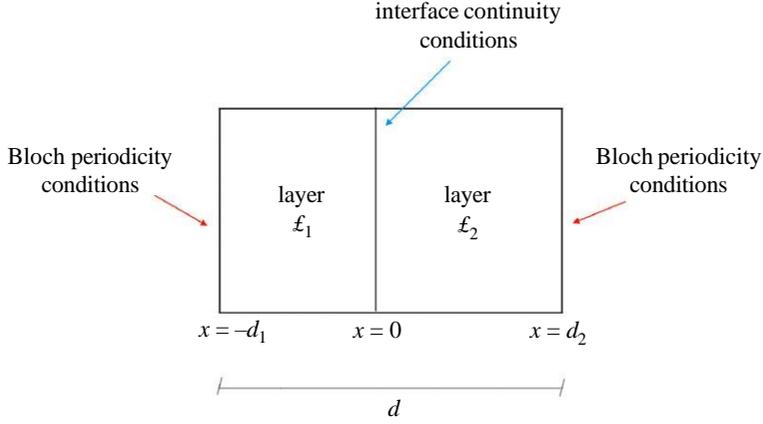

**Figure 4.** Wave propagation in layered media: unit cell and conditions imposed.

Substituting equations (3.3) and (3.5) into equations (3.6) and (3.7) provides the homogeneous system

$$\begin{bmatrix} 1 & -1 & 1 & -1 \\ E_1 k_1 & -E_2 k_2 & -E_1 k_1 & E_2 k_2 \\ e^{i(qd-k_1 d_1)} & -e^{i(k_2 d_2)} & e^{i(qd+k_1 d_1)} & -e^{-i(k_2 d_2)} \\ E_1 k_1 e^{i(qd-k_1 d_1)} & -E_2 k_2 e^{i(k_2 d_2)} & -E_1 k_1 e^{i(qd+k_1 d_1)} & E_2 k_2 e^{-i(k_2 d_2)} \end{bmatrix} \begin{bmatrix} A_1 \\ A_2 \\ B_1 \\ B_2 \end{bmatrix} = \begin{bmatrix} 0 \\ 0 \\ 0 \\ 0 \end{bmatrix} \quad (3.8)$$

whose non-trivial solution, obtained by imposing that the determinant of the matrix of coefficients vanishes

$$\det \begin{bmatrix} 1 & -1 & 1 & -1 \\ E_1 k_1 & -E_2 k_2 & -E_1 k_1 & E_2 k_2 \\ e^{i(qd-k_1 d_1)} & -e^{i(k_2 d_2)} & e^{i(qd+k_1 d_1)} & -e^{-i(k_2 d_2)} \\ E_1 k_1 e^{i(qd-k_1 d_1)} & -E_2 k_2 e^{i(k_2 d_2)} & -E_1 k_1 e^{i(qd+k_1 d_1)} & E_2 k_2 e^{-i(k_2 d_2)} \end{bmatrix} = 0, \quad (3.9)$$

yields the dispersion equation of the investigated layered system:

$$\cos(qd) - \cos(\omega \gamma_1)\cos(\omega \gamma_2) + \frac{1+\chi^2}{2\chi} \sin(\omega \gamma_1)\sin(\omega \gamma_2) = 0, \quad (3.10)$$

where, for conciseness,

$$\gamma_1 := \frac{d_1}{\sqrt{E_1/\rho_1}}, \quad \gamma_2 := \frac{d_2}{\sqrt{E_2/\rho_2}} \quad \text{and} \quad \chi := \frac{E_2 k_2}{E_1 k_1}. \quad (3.11)$$

In the low-frequency/long-wavelength approximation, i.e. when $\omega\gamma_1$ and $\omega\gamma_2$ are small [34], equation (3.10) takes the form

$$1 - \cos(qd) - \frac{\omega^2}{2}\left(\gamma_2^2 + \gamma_1^2 + \frac{(1+\chi^2)\gamma_1\gamma_2}{\chi}\right)$$
$$+ \frac{\omega^4}{4}\left(\gamma_1^2\gamma_2^2 + \frac{1+\chi^2}{3\chi}\left(\gamma_1\gamma_2^3 + \gamma_1^3\gamma_2\right)\right) - \frac{\omega^6}{72}\gamma_1^3\gamma_2^3\frac{(1+\chi^2)}{\chi} = 0. \quad (3.12)$$

The derived dispersion equation establishes a relationship between the frequency $\omega$ and the Bloch parameter $q$, allowing us to construct a complete band/dispersion diagram including all possible propagating and evanescent wave modes. As will be seen in the next section, two different approaches exist: $\omega(q)$, obtaining $\omega$ for a given real value of $q$, and $q(\omega)$, giving $q$ for specified real values of $\omega$ [35–37].

It should be noted that, in general, the in-plane elastic wave propagation involves a complicated coupling between P-waves (compressional waves) and SV-waves (vertically polarized shear waves). This can be observed by considering the simple example of in-plane elastic waves propagating in two-dimensional periodic multilayered media constituted by alternating a number of isotropic layers. By virtue of the Floquet theorem for periodic structures, the dispersion characteristics of the whole structure can be obtained by focusing on a single unit cell composed of a generic number of $n$ different layers.

Writing the corresponding time-harmonic elastodynamic problem leads to

$$\left.\begin{array}{l}\frac{\partial}{\partial x}\left((2\mu_j + \lambda_j)\frac{\partial u_j}{\partial x} + \lambda_j \frac{\partial v_j}{\partial y}\right) + \frac{\partial}{\partial y}\left(\mu_j\left(\frac{\partial u_j}{\partial y} + \frac{\partial v_j}{\partial x}\right)\right) = \rho_j \frac{\partial^2 u_j}{\partial t^2} \\ \frac{\partial}{\partial y}\left((2\mu_j + \lambda_j)\frac{\partial v_j}{\partial y} + \lambda_j \frac{\partial u_j}{\partial x}\right) + \frac{\partial}{\partial x}\left(\mu_j\left(\frac{\partial v_j}{\partial x} + \frac{\partial u_j}{\partial y}\right)\right) = \rho_j \frac{\partial^2 v_j}{\partial t^2}\end{array}\right\} \quad (3.13)$$

and

with $u_j = u_j(x, y, t)$ and $v_j = v_j(x, y, t)$ denoting, respectively, the longitudinal and transverse displacement of the $j$th layer, $\mu_j$ and $\lambda_j$ the Lamé constants characterizing the $j$th layer, $\rho_j$ its density. We thus obtain a complex system of coupled equations to which the periodicity and continuity conditions of displacements and stresses at the layers interfaces need to be added.

Even in the simplest case of a unit cell composed by only two layers, this results in a dispersion equation given by the determinant of a $8 \times 8$ matrix that, in general, can only be evaluated numerically [38]. Only in very simple cases such as periodic bi-layered media composed of isotropic elastic layers, the $8 \times 8$ determinant can be analytically expanded and, after very lengthy and cumbersome algebraic manipulations, a closed-form dispersion relation can be obtained [39]. The derived equation, however, is very complicated and difficult to use in a practical context because of the coupling between $P$- and SV-waves both in the same layer and between two different layers [39]. A significant simplification is possible for small thickness ratios between the two repeating layers since, in this case, the dispersion equation decouples, and two independent dispersion relations can be obtained (one for the $P$-waves and one for the SV-waves) [39]. A further simplification is provided by considering in-plane waves propagating along the thickness direction, i.e. direction $x$ in figure 1. Also in this case, the in-plane waves can be decoupled into pure longitudinal ($P$) and pure transverse shear (SV) waves, so that the two wave propagation modes can be studied separately. This simplifies the problem, when compared with solving the coupled counterpart [40].

This simplification does not apply when viscoelastic multilayered composites are involved. Because of the intricate wave attenuation characteristics at the layer interfaces, in this case the analytical approaches and derived considerations developed for elastic multilayered composites cannot be applied directly [41]. However, in the case of infinitely periodic multilayered composites composed of alternating viscoelastic and elastic layers, as in the problem treated here, Tahidul Haque *et al.* [41] found that the attenuation (and corresponding band gaps) of harmonic in-plane waves propagating along different orientations only occurs in the direction perpendicular to the layer interfaces while no band gaps appear for oblique incidence, regardless of the incident angle. This is related to the behaviour of the viscoelastic layer for which, when attached to an elastic one, wave attenuation exists only for perpendicular incidence. It also emerged in [41] that the analytical techniques employed in elastic multilayered composites, e.g. the transfer matrix method, can be applied to investigate the wave attenuation characteristics of alternating elastic/viscoelastic multilayer composites and, in particular, the decoupled $P$- and SV-waves modes can be separately analysed and the corresponding dispersion equation obtained. In this paper, for simplicity, we have chosen to focus only on longitudinal $P$-waves. This simplification reduces the algebraic complexity of the problem, allowing us to derive a closed-form dispersion relation that clearly demonstrates the phenomenology we seek to address, i.e. wave dissipation in layered media with suture joints, and more easily enables the influence of the essential physical attributes to be identified.

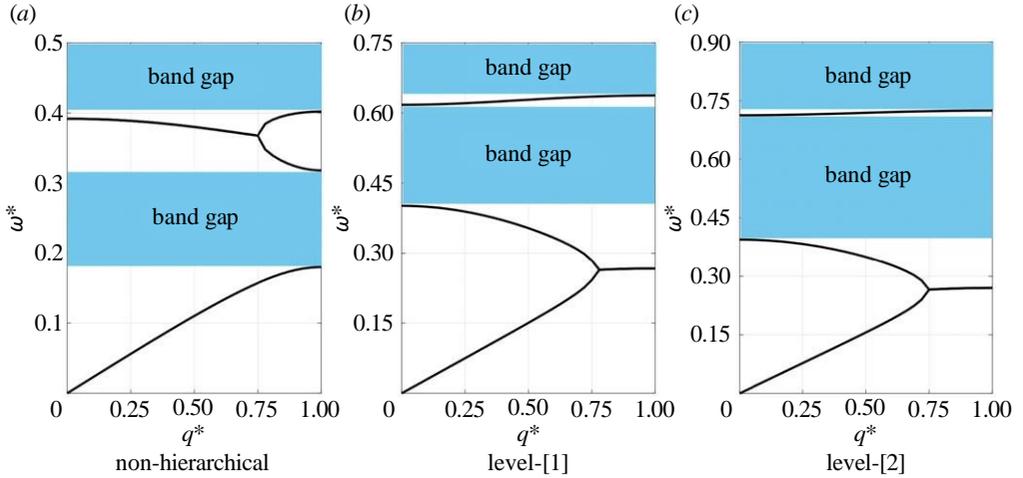

**Figure 5.** Non-dissipative linear elastic medium: dispersion diagrams for triangular sutures in the case of a (*a*) non-hierarchical and (*b*,*c*) hierarchical configuration with one (*b*) and two (*c*) levels of hierarchy. Focus on the first bandgap.

## 4. Results and discussion

According to the above formulation, this section aims to understand how the suture characteristics, i.e. hierarchical or non-hierarchical configuration, geometric profile, mechanical properties of the two constituents, affect the dispersion curves and frequency bandgaps of the examined layered medium. Two different scenarios are investigated: a non-dissipative linear elastic behaviour of the system and a dissipative viscoelastic one. In both cases, the overall layered
medium is composed by periodically alternating a layer $\ell_1$ with Young's modulus $E_1 = 6$ GPa, shear modulus $G_1 = 2.3$ GPa and density $\rho_1 = 2000$ kg m$^{-3}$ (properties of bone, treated as an isotropic material [20]), and a layer $\ell_2$ whose hard and soft constituent phases have Young's
modulus, shear modulus and density given, respectively, by $E_T = E_1$, $G_T = G_1$, $\rho_T = \rho_1$ and $E_L = 10^{-2}E_T$, $G_L = 10^{-2}G_T$, $\rho_L = 10^{-2}\rho_T$. Also, a teeth volume fraction of $\varphi_T = 0.9$ and a tooth tip angle $\beta_S = 30°$ have been considered (see §2a). Finally, we have assumed that the thickness of the layer $\ell_2$ is $d_2 = 0.1d_1$ and, for simplicity, that the total thickness $d$ of the unit cell, i.e. $\ell_1 + \ell_2$, has a unit value (figures 1 and 3).

### (a) Non-dissipative linear elastic layered medium

A linear elastic response of the bi-layered system $\ell_1 + \ell_2$ is obtained by assuming $\xi = \eta = 0$, so that both $\ell_1$ and the two constituent phases of $\ell_2$, the teeth and the interface layer, have a linear elastic behaviour.

In the linear elastic case, where the frequencies, $\omega$, and the wavevectors, $q$, are real-valued, the dynamic response can be completely captured by the $\omega(q)$ approach. That is, by solving for $\omega$ the dispersion relation in equation (3.12) for discrete values of $q$, taken in a specific range. The corresponding dispersion diagrams are illustrated in figure 5 for the case of a triangular suture profile, as a function of the non-dimensional frequency $\omega^* := \omega d/\sqrt{E_1/\rho_1}$ and wavenumber $q^* := qd/\pi \in [0, 1]$. Both the non-hierarchical (figure 5*a*) and hierarchical configurations with one (figure 5*b*) and two (figure 5*c*) orders of hierarchy are reported. For simplicity, in the hierarchical case, self-similar conditions [42] are assumed, together with the hypothesis of a constant value of the teeth volume fraction at all levels.

The investigated geometry exhibits two bandgaps, within which wave propagation is forbidden. Specifically, if we focus on the first bandgap, for the non-hierarchical system it ranges between $\omega^* = 0.17$ and $\omega^* = 0.32$ (figure 5*a*). Direct comparison of figure 5*a* and 5*b*,*c* reveals two important advantages of adding hierarchy to the starting layered system. Firstly, due to the self-similar configuration at all levels, the trend of the dispersion curves is conserved, together with the appearance of two bandgaps. Secondly, if compared to the non-hierarchical configuration, the bandgap at low frequencies not only experiences a small shift to higher values of $\omega^*$, but also becomes significantly wider. To make it more clear, let us focus on figure 5b. It emerges that
adding one level of hierarchy leads to a lower bandgap located at $\omega^* = [0.40, 0.62]$, that is 57%
wider than that in the non-hierarchical case. Additional improvements are obtained with two levels of hierarchy (figure 5c), being the bandgap at lower frequencies, located at $\omega^* = [0.39, 0.71]$, approximately 45% wider than that in the first-level structure.

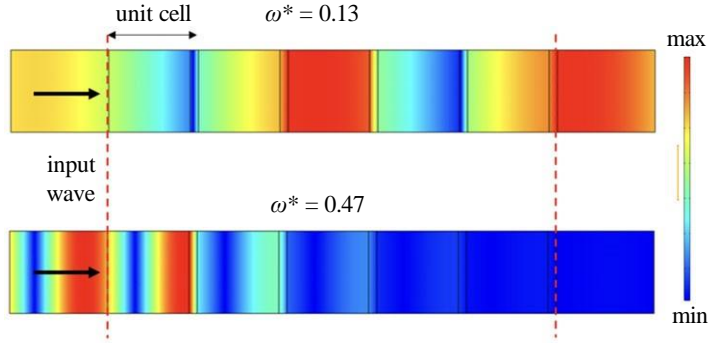

**Figure 6.** Frequency domain FEM simulations of wave propagation in a lattice composed of four non-hierarchical unit cells, indicating that propagation is not inhibited for frequencies below the calculated band gaps (upper figure), while filtering occurs when frequencies fall in one of the band gaps (lower figure). Colour scale indicates displacement amplitudes from blue (minimum) to red (maximum).

To confirm these results, finite-element method (FEM) simulations are performed using COMSOL Multiphysics. The first, non-hierarchical structure is considered as a representative example. A sample consisting of a lattice of five unit cells is considered, and frequency domain simulations are performed by exciting plane waves at one end of the sample, in the direction of the unit cell array. Low-reflecting boundary conditions are implemented at the other end, while symmetry conditions are enforced at upper and lower boundaries. Results, shown in figure 6, indicate that when the exciting frequency falls outside the band gaps calculated in figure 5a, waves can propagate freely through the lattice (upper figure); instead, when the frequency falls within one of the band gaps (lower figure), the wave amplitude is rapidly damped and falls to zero in the space of a couple of unit cells.

Consider now figure 7, where the comparison between the bandgaps range of the triangular suture profile is compared to that of different suture geometries: trapezoidal, anti-trapezoidal and rectangular. They are obtained by different values of the ratio $\alpha_S/\beta_S$ (figure 2): $\alpha_S/\beta_S = 0.2$
for the trapezoidal, $\alpha_S/\beta_S = -0.6$ for the anti-trapezoidal, $\alpha_S/\beta_S = 0$ for the rectangular. In the
non-hierarchical case, similarly to the triangular profile, all the geometries exhibit a first bandgap that approximately spans the range $\omega^* = [0.18, 0.3]$. However, in terms of bandgap amplitude, the
triangular geometry displays the highest value, 0.16, which is, respectively, 30%, 43% and 30% larger than that of the trapezoidal (0.12), anti-trapezoidal (0.11) and rectangular (0.12). For all the examined geometries, figure 7 also reveals that adding hierarchy leads to an improvement in the bandgap amplitude even if for the trapezoidal, anti-trapezoidal and rectangular sutures the benefits are more evident, if compared with the triangular one. In particular, it becomes 57%, 89%, 112%, 83% wider, respectively, in the case of triangular, trapezoidal, anti-trapezoidal and rectangular geometry with one level of hierarchy. Adding a second level of hierarchy yields an additional improvement in the bandgap amplitude also for the trapezoidal, anti-trapezoidal and rectangular profile. For these geometries, the improvement is, respectively, of approximately 56%, 51%, 57%, if compared to the first-level structure. Values that, again, are larger than the 45% of the triangle. The consequence of these larger improvements provided by the introduction of hierarchy can be observed in figure 7 by comparing the bandgap amplitude of the four considered geometries. If compared to the triangular suture, it emerges that the trapezoidal, anti-trapezoidal and rectangular ones provide a bandgap approximately 4.5% and 9% wider, respectively, with one and two levels of hierarchy.

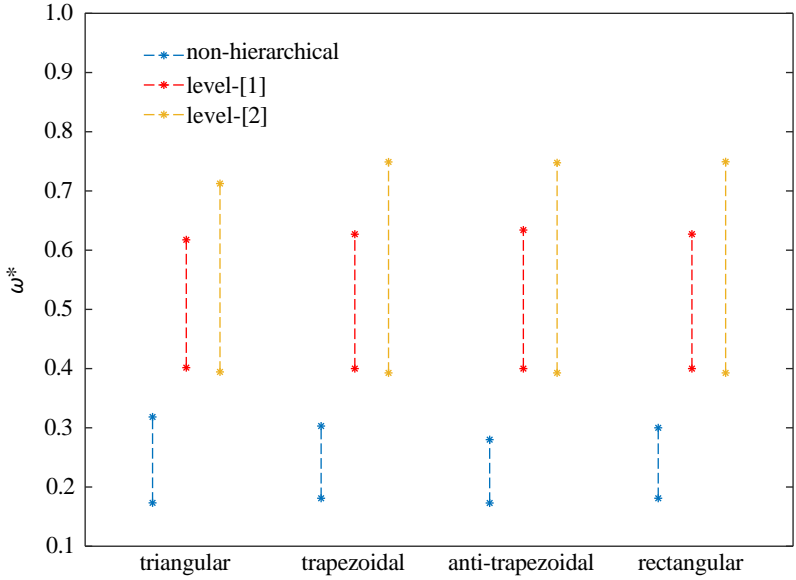

**Figure 7.** Non-dissipative linear elastic medium: comparison of the bandgaps amplitude corresponding to different suture geometries in the non-hierarchical and hierarchical cases. Focus on the first bandgap.

It should be noted that the above considerations are valid not only for the specific selected values of the parameter $\alpha_S/\beta_S$ (respectively, 0.2 and −0.6 for the trapezoidal and anti-trapezoidal profile), but can be extended to the general case of trapezoidal and anti-trapezoidal geometries, described by different values of $\alpha_S/\beta_S$. This can be seen in figure 8, where the first bandgap amplitude corresponding to the non-hierarchical and hierarchical configurations is plotted versus the parameter $\alpha_S/\beta_S$. In this case, too, it emerges that the benefits of adding hierarchy are more evident for the trapezoidal ($0 < \alpha_S/\beta_S < 1$), anti-trapezoidal ($-1 < \alpha_S/\beta_S < 0$) and rectangular geometries, if compared to the triangular one.

## (b) Dissipative viscoelastic layered medium

Viscoelastic dissipation is taken into account for $\xi \neq 0$ and $\eta \neq 0$. This condition coincides with a viscoelastic mechanical response of the interface layer in $\pounds_2$ that, in this case, has Young's modulus and shear modulus given, respectively, by $E_L + i\omega\xi$ and $G_L + i\omega\eta$. These relations, inserted into equations (2.2) and (2.3), lead to frequency-dependent complex effective properties of the layer $\pounds_2$ and, consequently, to a complex-valued dispersion relation.

When considering viscoelasticity, all the waves become attenuating or evanescent and, in contrast to the linear elastic case, they are described by complex-valued frequencies or complex-valued wavenumbers, with imaginary parts characterizing, respectively, the temporal or spatial wave attenuation. Two different approaches can be used to investigate the dynamic response of viscoelastic media and obtain the corresponding dispersion diagram. The first, known as the $\omega(q)$ approach, is commonly adopted in impulsive phenomena and consists in solving the dispersion relation by assuming freely propagating waves with complex frequencies and real wavenumbers [43]. The second, useful for steady-state wave fields, is known as the $q(\omega)$ approach and solves the dispersion relation by considering harmonic waves with real frequencies and complex wavevectors [43].

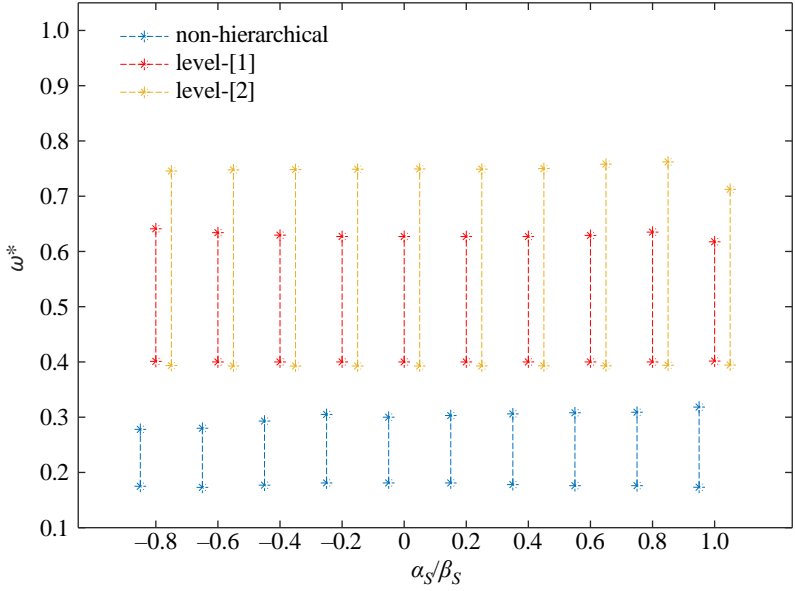

**Figure 8.** Non-dissipative linear elastic medium: comparison of the first bandgap amplitude corresponding to different values of $\alpha_S/\beta_S$ in the non-hierarchical and hierarchical cases.

In Dynamics, when dealing with viscoelastic materials, it is convenient to use the elastic–viscoelastic correspondence principle [30], according to which the solution techniques developed for the linear elastic framework can be directly used for viscoelastic media simply by replacing the elastic material parameters with their frequency-dependent complex counterparts. Given that the results presented in §3 still apply, the dispersion diagrams can be derived from equation (3.12) by applying, in this case, the $q(\omega)$ approach. This consists of solving for the complex values of $q$ the dispersion relation for discrete real values of $\omega$, selected within a range of interest.

Results are presented in figure 9 for the case of a non-hierarchical triangular suture profile and $\xi = \eta = 5$ Pa s, which is a typical value for polymeric materials. In accordance with §4a, non-dimensional frequencies, $\omega^* := \omega d/\sqrt{E_1/\rho_1}$ and wavenumbers, $q^* := qd/\pi$, are used. For completeness, both the real, $\text{Re}(q^*)$, and imaginary, $\text{Im}(q^*)$, part of the wavevector are plotted, the first describing the harmonic wave propagation and the second representing the wave spatial attenuation. The parameter $\delta^* := 2\text{Im}(q^*)/\text{Re}(q^*)$, which characterizes the wave attenuation level, is also plotted against the frequency $\omega^*$, providing the so-called attenuation spectrum.

Comparing the dispersion diagram in figure 9a with those obtained in the linear elastic case (figure 5), it emerges that this level of viscoelasticity does not influence the general trend of the dispersion curves, including bandgaps. As expected, figure 9b shows a non-zero imaginary part of the wavevector in correspondence with the bandgap, revealing a wave attenuation that, in accordance with the adopted Kelvin–Voigt model (cf. §2), increases linearly with frequency. A curve-rounding effect [44] occurring where the two bandgaps are located is also revealed from figure 9b.

In terms of the parameter $\delta^*$, figure 10 illustrates that hierarchy is advantageous to increase attenuation for the considered system. By focusing on the first bandgap, it emerges that one level of hierarchy provides an enhancement of the maximum attainable value of $\delta^*$ of approximately 130%. A less significant increase in the level of wave attenuation is obtained by adding a second level of hierarchy, with an improvement in the maximum value of $\delta^*$ of approximately 34% if compared to the first level system.

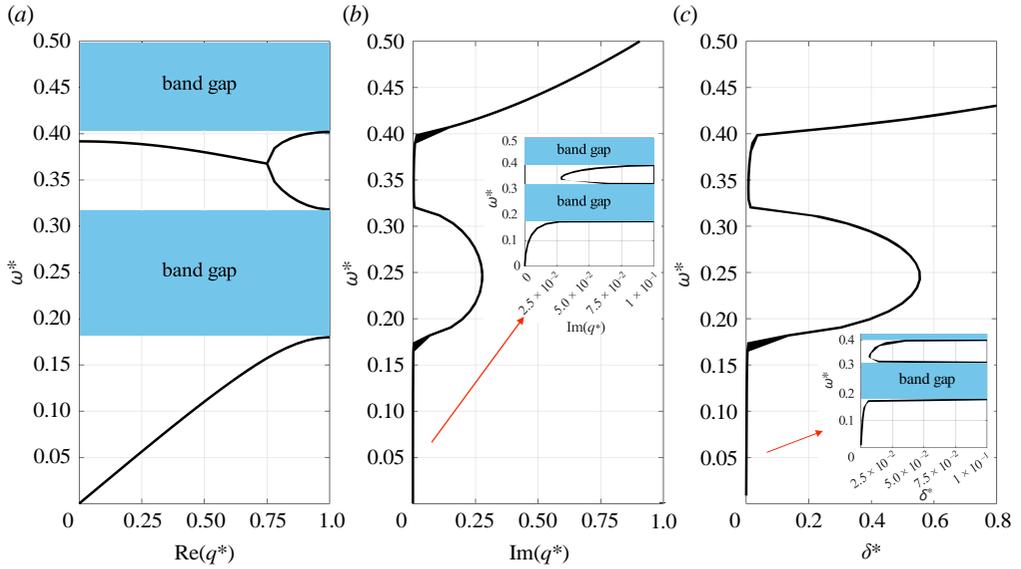

**Figure 9.** Dissipative viscoelastic medium: dispersion diagram and attenuation spectrum for triangular sutures in the case of a non-hierarchical configuration.

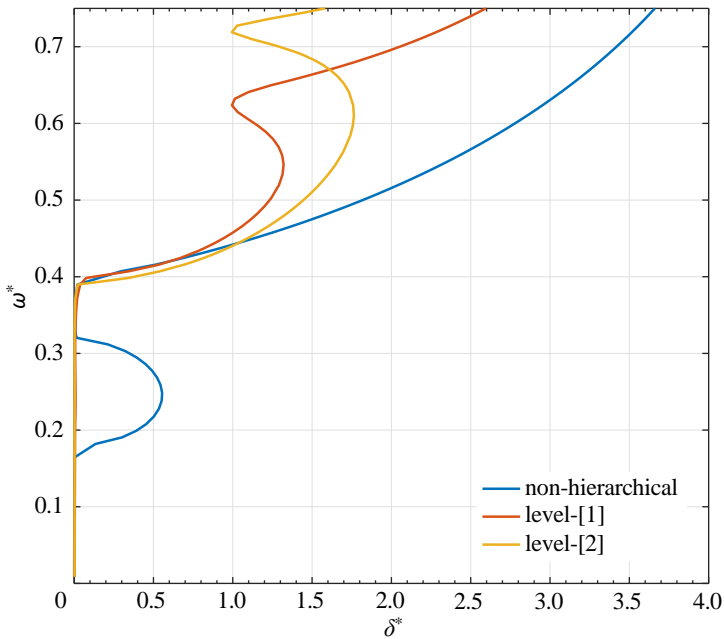

**Figure 10.** Comparison of the wave attenuation level for triangular sutures in the non-hierarchical and hierarchical cases (level- [1] and level-[2]) with focus on the first bandgap.

Some peculiarities emerge if we consider different suture geometries: trapezoidal, anti- trapezoidal and rectangular, corresponding, respectively, to $\alpha_S/\beta_S = 0.2$, $\alpha_S/\beta_S = -0.6$, $\alpha_S/\beta_S = 0$. The first can be observed in table 1, where the maximum attainable value of $\delta^*$, $\delta^*_{max}$, of the triangular profile is compared to that of the other geometries. In the non-hierarchical case, it
appears that the triangular suture provides the maximum level of wave attenuation, 0.56, which is approximately 3% higher than that of the other geometries. Conversely, when we add one level of hierarchy, the triangular suture is slightly less attenuating than the trapezoidal and rectangular geometries (approx. 1.5%). This consideration still applies with two levels of hierarchy, being the triangular profile approximately 5% less attenuating than the trapezoidal and rectangular ones. Among the geometries considered, we can thus say that the triangular is the optimum for obtaining the largest level of wave attenuation only in the non-hierarchical case. When hierarchy is introduced, superior performances are provided by trapezoidal and rectangular sutures. The second peculiarity is related with the introduction of hierarchy, as summarized in figure 11. Generally, it emerges that hierarchy improves the parameter $\delta^*$, independently of the considered geometry. The improvement, in particular, is of averagely 140–150% with one level of hierarchy and of 35–40% with two levels. The largest benefits are observed for the trapezoidal and rectangular profiles while the triangular geometry experiences the lowest improvement.

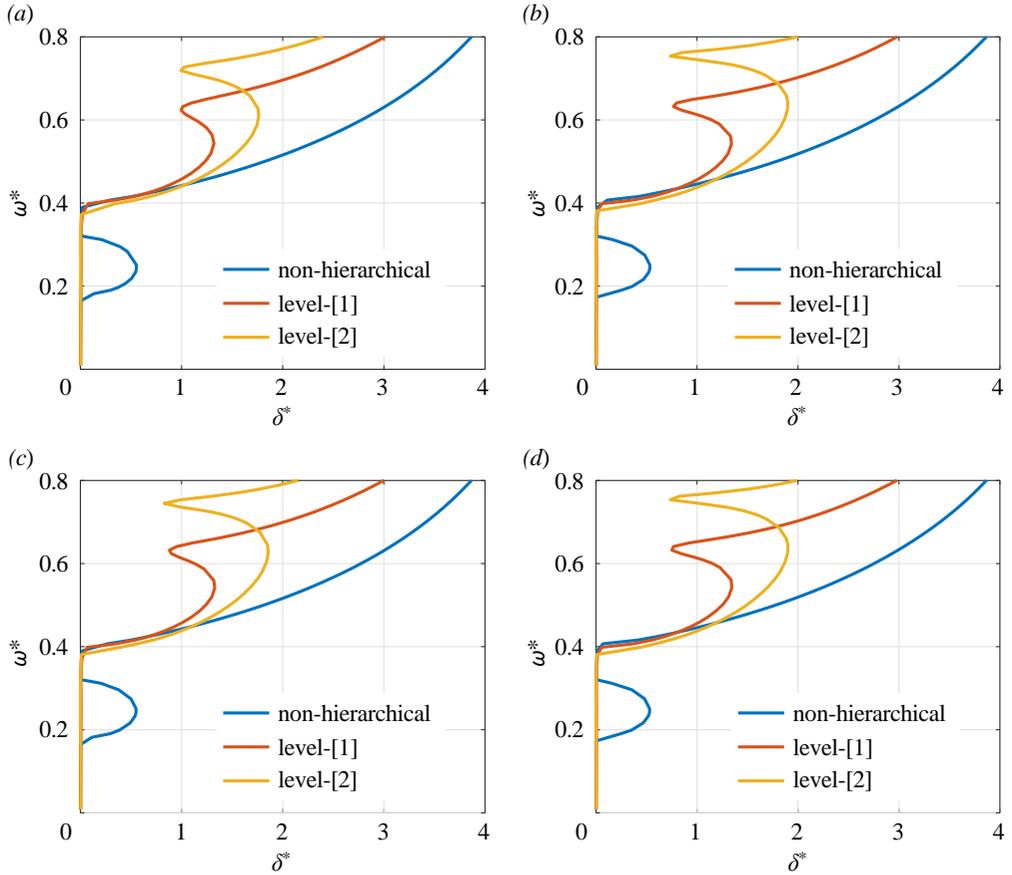

**Figure 11.** Comparison of the improvement in the wave attenuation level obtained by adding hierarchy in the case of (*a*) triangular, (*b*) trapezoidal, (*c*) anti-trapezoidal and (*d*) rectangular suture profile. Focus on the first bandgap.

In the context of suture joints, these results reveal that there is a synergy of hierarchy and geometric features in obtaining superior levels of wave attenuation.

As a final observation, it is important to note that while in figures 9–11 the imaginary part of the wavenumber assumes small values, it does not vanish. In particular, as highlighted in figure 9*b*,*c*,
the imaginary part and corresponding parameter $\delta^*$ are non-zero for frequencies within the first
two bandgaps and for frequencies below the first bandgap. This consideration also applies for figures 10 and 11.

**Table 1.** Comparison of the maximum attainable value of wave attenuation corresponding to different suture geometries.

|  | $\delta_{m\ ax}^{*}$ | | |
|---|---|---|---|
|  | non-hierarchical | level-[1] | level-[2] |
| triangular | 0.56 | 1.32 | 1.77 |
| trapezoidal | 0.53 | 1.33 | 1.87 |
| anti-trapezoidal | 0.54 | 1.32 | 1.85 |
| rectangular | 0.53 | 1.34 | 1.88 |

# 5. Conclusion

We have investigated the dynamic response of periodic layered media with suture joints in a configuration that resembles the common millet seedcoat. Namely, a layered medium obtained by periodically alternating a first homogeneous layer and a second suture layer where two interdigitating stiff phases are joined by a thin compliant interface layer. The general case of a trapezoidal suture profile has been examined in the presence of two fundamental ingredients: hierarchy and viscoelasticity.

First, based on the elastic–viscoelastic correspondence principle, closed-form expressions for the effective stiffness and mass density of the suture layer have been found for both the non- hierarchical and hierarchical arrangement. These relations, closely related to the suture geometric and mechanical properties, were then inserted into the governing equation of motion of the examined layered medium whose solution, in terms of displacements and stress field, was obtained by assuming time-harmonic waves and infinitesimal deformations. The application of the Spectral Element Method in conjunction with the Bloch theorem, commonly used for problems involving time-harmonic waves travelling within periodic media, provide the dispersion relation of the investigated system that, once solved, leads to a complete band diagram including propagating and evanescent dispersion modes. From this point of view, a strong influence of the suture morphology emerges, i.e. geometric profile, hierarchical or non- hierarchical configuration, and mechanical properties, i.e. linear elastic or viscoelastic response. To examine more closely this dependence, we have illustrated the example of a self-similar hierarchical medium with two levels of hierarchy.

The analysis reveals that, if compared to the non-hierarchical counterpart, introducing hierarchy leads to wider bandgaps and, simultaneously, to a shift of the bandgaps to higher frequencies. Also, when viscoelasticity is taken into account, adding hierarchy provides an enhancement of the wave attenuation level. The second finding of our study is that there is a synergy of hierarchy and geometric features in obtaining superior levels of wave attenuation.

More generally, it is known that the design and mechanical properties of suture joints, e.g. in the human skull, are optimized for impact damping. This design helps to reduce the peak stress and strain levels that occur during an impact. The present work also shows that by locally modifying equivalent (homogenized) mechanical properties, these structures can alter the behaviour of waves propagating across sutures, at least in the low-frequency domain, generating band gaps and strong damping at specific frequencies.

There are several bioinspired artificial applications that mimic the structure and function of suture joints, to exploit their functionality. One example is the development of interlocking joint structures in materials used for impact protection and energy absorption. These structures are designed to mimic the complex interlocking structure of the suture joints in the skull or in other biological examples, which helps to distribute and absorb impact energy. Another example is the use of soft interfaces between stiff materials, similar to the suture joints in the skull. These soft interfaces can be used to absorb and dissipate impact energy in a similar way to the suture joints, and are used in a range of applications, including sports equipment, automotive components and protective gear. The analysis presented herein can extend the functionality of these applications, explicitly accounting for vibration damping, with the possibility of tuning the band gap ranges to the frequencies of modal vibrations, to more efficiently attenuate them.

In conclusion, this comprehensive study, the first to explore the dynamic behaviour of a new class of viscoelastic hierarchical layered media with suture joints, may be relevant to the development of bioinspired devices with tailorable wave attenuation properties via structural hierarchy. Compared to previous works, the additional consideration of viscoelasticity can increase the reliability of the analysis for a possible better understanding of biological systems with suture joints.

Data accessibility. This article has no additional data.
Authors' contributions. F.O.: conceptualization, formal analysis, investigation, writing—original draft; F.B.: investigation, validation, writing—review and editing; N.M.P.: conceptualization, funding acquisition, project administration, resources, supervision, writing—review and editing.
All authors gave final approval for publication and agreed to be held accountable for the work performed therein.
Conflict of interest declaration. We declare we have no competing interests.
Funding. The authors are supported by the EU H2020 FET-Open project Boheme (grant no. 863179).

# Appendix A. Homogeneization

Let us focus on the trapezoidal suture joint illustrated in figure 12. In line with §2a, let us denote with $E_2$ and $G_2$, respectively, its effective Young's modulus in the longitudinal direction $e_1$ and effective shear modulus.

Closed-form relations for $E_2$ and $G_2$ accounting the viscoelastic effect can be derived in the frequency domain by alternately prescribing a periodic loading condition

$$\hat{\sigma}_1 = \begin{bmatrix} \hat{\sigma}_{11} \neq 0 \\ \hat{\sigma}_{22} = 0 \\ \hat{\sigma}_{12} = 0 \end{bmatrix} \quad \text{and} \quad \hat{\sigma}_2 = \begin{bmatrix} \hat{\sigma}_{11} = 0 \\ \hat{\sigma}_{22} = 0 \\ \hat{\sigma}_{12} \neq 0 \end{bmatrix}, \qquad (A\,1)$$

described by

$$\hat{\sigma}_{ij} = \Gamma_{ij} e^{i\omega t}, \qquad i, j = 1, 2, \qquad (A\,2)$$

with $\Gamma_{ij}$ the amplitude and $\omega$ the circular frequency of a time-harmonic plane wave propagating along the medium. According to the elastic–viscoelastic correspondence principle [30], the solution techniques developed for the linear elastic framework can be directly applied for viscoelastic media simply by replacing the elastic material parameters with their frequency- dependent complex counterparts. For the examined viscoelastic interface layer, they are given by

$$\tilde{G}_L \equiv G_L + i\omega\eta \quad \text{and} \quad \tilde{E}_L \equiv E_L + i\omega\xi, \qquad (A\,3)$$

with $G_L$ and $E_L$, respectively, the layer shear modulus and Young's modulus, $\eta$ and $\xi$ the viscosity coefficients. A classical mechanics-based analysis [7,8] can thus be applied, providing, in the loading condition $\hat{\sigma}_1$, the complex effective Young's modulus in the longitudinal direction

$$E_2 = \left[ \frac{\lambda_S - 2h_L}{\lambda_S E_T} + \frac{2h_L}{\lambda_S}\left( \frac{\cos^2\alpha_S \sin^2\alpha_S}{\tilde{G}_L} + \frac{\cos^4\alpha_S}{\tilde{E}_L} \right) \right]^{-1}, \qquad (A\,4)$$

with $E_T$ Young's modulus of the teeth, $\lambda_S$ the wavelength of the suture, $h_L$ the thickness of the interface layer, $\alpha_S$ the angle defining the suture geometry (figure 12).

Similarly, the loading state $\hat{\sigma}_2$ leads to the complex effective shear modulus

$$G_2 = \frac{(\lambda_S - 2h_L)\Phi_1}{((\Phi_1/4\Phi_2)(3 + (5\Phi_2/4E_T \tan^2\beta_S)) + (\lambda_S \tan\alpha_S/2 A_S)((2 h_L/\lambda_S \psi_{\alpha\beta}) + (3\tilde{G}_L/2\Phi_2)(1 + (3\Phi_2/4 E_T \tan^2\beta_S))))\lambda_S}, \qquad (A\,5)$$

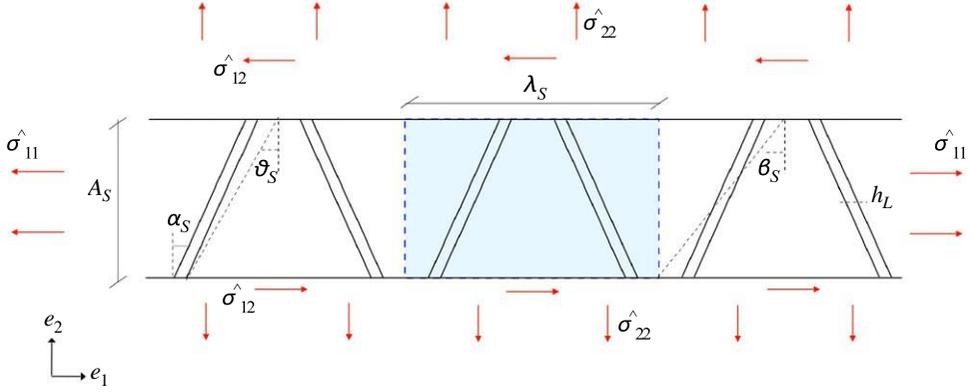

**Figure 12.** The trapezoidal suture joint subjected to the loading conditions $\hat{\sigma}_1$ and $\hat{\sigma}_2$.

where $A_S$ is the amplitude of the suture waveform, $\beta_S$ the angle defined in figure 12 and, to simplify the notation,

$$\psi_{\alpha\beta} := \frac{\tan \alpha_S}{\tan \beta_S}, \quad \Phi_1 := \tilde{G}_L \tan^2 \alpha_S + \tilde{E}_L \quad \text{and} \quad \Phi_2 := \frac{G_T(\lambda_S - 2h_L) + 2\tilde{G}_L h_L}{\lambda_S}. \tag{A 6}$$

# Appendix B. Comparison between elastic and electromagnetic wave propagation

In the context of metamaterials, the homogenization approach adopted in this paper is a well-accepted mathematical tool because of its ability to significantly reduce the complexity of modelling such periodically heterogeneous structures. Here, in particular, the method has been used to investigate the elastic wave dispersion in layered structures. However, different phenomena can be analysed [45].

An interesting example is the case of the electromagnetic waves propagation that, from a mathematical point of view, shares certain common features with the elastic counterpart considered here. Although from a physical perspective they are quite different, being the electromagnetic waves able to travel through vacuum, both type of waves can be described by the same governing equation. By restricting the analysis to the two-dimensional $x$–$y$ plane, the latter, in its general form, can be expressed by [46]:

$$\nabla \left( \frac{1}{p_1(\mathbf{r})} \nabla \Gamma(\mathbf{r}, t) \right) - p_2(\mathbf{r}) \frac{\partial^2 \Gamma(\mathbf{r}, t)}{\partial t^2} = 0 \tag{B 1}$$

being $\nabla(\cdot) := (\partial(\cdot)/\partial x, \partial(\cdot)/\partial y)$, $\mathbf{r} = (x, y)$ the position vector, $\Gamma(\mathbf{r}, t)$ a scalar function describing a particular physical field, $p_1(\mathbf{r})$ and $p_2(\mathbf{r})$ two suitable parameters of the medium such that $1/p_1(\mathbf{r})p_2(\mathbf{r})$ defines the phase velocity of the wave. By considering a time-harmonic plane wave propagating along the medium, given by

$$\Gamma(\mathbf{r}, t) = \tilde{\Gamma}(\mathbf{r}) e^{-i\omega t} \tag{B 2}$$

with $\tilde{\Gamma}(\mathbf{r})$ the wave amplitude and $\omega$ the wave frequency, and substituting equation (B 2) into equation (B 1), lead to the final expression

$$\nabla \left( \frac{1}{p_1(\mathbf{r})} \nabla \tilde{\Gamma}(\mathbf{r}) \right) + \omega^2 p_2(\mathbf{r}) \tilde{\Gamma}(\mathbf{r}) = 0. \tag{B 3}$$

The solution of equation (B 3), obtained by applying suitable boundary conditions, provides the function $\tilde{\Gamma}(\mathbf{r})$ so that, from equation (B 2), the investigated physical field $\Gamma(\mathbf{r}, t)$ (and derived physical quantities) can be obtained.

Let us firstly consider the electromagnetic waves, i.e. a coupled oscillating electrical field $E(r)$ and magnetic field $H(r)$ always perpendicular to each other. From a physical point of view, a time-harmonic electromagnetic wave with frequency $\omega$ propagating along a medium with dielectric permittivity $\xi(r)$ and magnetic permeability $\mu(r)$, is described by the Maxwell equations [47]

$$\nabla \times E(r) = -i\omega\mu(r)H(r)$$
$$\nabla \times H(r) = i\omega\xi(r)E(r) \tag{B 4}$$

and

with $i := \sqrt{-1}$. If we assume, for simplicity, that $\mu(r)$ and $\xi(r)$ are independent of $r$, i.e. an isotropic medium with $\mu(r) \equiv \mu$ and $\xi(r) \equiv \xi$, and we solve equation (B 4) for either $E(r)$ or $H(r)$, we obtain the following two forms of electromagnetic wave propagation equation:

$$\nabla \left(\frac{1}{\mu}\nabla E(r)\right) + \omega^2 \xi E(r) = 0 \tag{B 5}$$

and

$$\nabla \left(\frac{1}{\xi}\nabla H(r)\right) + \omega^2 \mu H(r) = 0. \tag{B 6}$$

It can be easily verified that, if written in terms of transverse electric (TE) or transverse magnetic (TM) field, i.e. the two polarization modes for in-plane electromagnetic waves, equation (B 5) and (B 6) coincide with equation (B 3).

In TE polarization, in particular, the electric field has only one non-zero component pointing in the out-of-plane direction (the $z$-direction in the present case), so that

$$E(r) = (0, 0, E_z(r)) \tag{B 7}$$

while, from equation (B 4), the magnetic field is given by

$$i\omega\mu H(r) = \nabla \times E(r) = \left(-\frac{\partial E_z(r)}{\partial y}, \frac{\partial E_z(r)}{\partial x}, 0\right). \tag{B 8}$$

Substituting equation (B 7) into equation (B 5), leads to the following wave propagation equation:

$$\frac{1}{\mu}\left(\frac{\partial^2 E_z(r)}{\partial x^2} + \frac{\partial^2 E_z(r)}{\partial y^2}\right) + \omega^2 \xi E_z(r) = 0 \tag{B 9}$$

that, if solved, suffices for the determination of both the electric and magnetic field. As it can be seen, the obtained expression exactly matches equation (B 3), provided that

$$(\tilde{\Gamma}(r), p_1(r), p_2(r)) \rightarrow (E_z(r), \mu(r) \equiv \mu, \xi(r) \equiv \xi). \tag{B 10}$$

Analogous considerations can be made for TM polarization, described by a magnetic field pointing in the out-of-plane direction $z$

$$H(r) = (0, 0, H_z(r)) \tag{B 11}$$

and an electric field expressed by

$$i\omega\xi E(r) = -\nabla \times H(r) = \left(\frac{\partial H_z(r)}{\partial y}, -\frac{\partial H_z(r)}{\partial x}, 0\right). \tag{B 12}$$

From equation (B 6), it follows

$$\frac{1}{\xi}\left(\frac{\partial^2 H_z(r)}{\partial x^2} + \frac{\partial^2 H_z(r)}{\partial y^2}\right) + \omega^2 \mu H_z(r) = 0 \tag{B 13}$$

that coincides with the general form of wave propagation equation (B 3), provided that

$$(\tilde{\Gamma}(r), p_1(r), p_2(r)) \rightarrow (H_z(r), \xi(r) \equiv \xi, \mu(r) \equiv \mu). \tag{B 14}$$

Also in this case, the solution of equation (B 13) suffices for the determination of both the magnetic and electric field.

Finally, let us consider the elastic waves propagation and, for simplicity, let us restrict our analysis to the case of time-harmonic vibrations. In particular, let us focus on anti-plane shear waves propagation that, for reasons of identical polarization perpendicular to the incidence plane, are closely analogous to the previously described transverse electromagnetic waves [46]. This can be seen by observing that for time-harmonic anti-plane shear waves of frequency $\omega$, the displacement field $\mathbf{u}(\mathbf{r})$ takes the form [48]

$$\mathbf{u}(\mathbf{r}) = (0, 0, u_z(\mathbf{r})). \tag{B 15}$$

It is thus a special state of out-of-plane mode of deformation and the corresponding wave propagation equation, written for an isotropic medium with shear modulus $G(\mathbf{r}) \equiv G$ and density $\rho(\mathbf{r}) \equiv \rho$, is given by [48]

$$G \left( \frac{\partial^2 u_z(\mathbf{r})}{\partial x^2} + \frac{\partial^2 u_z(\mathbf{r})}{\partial y^2} \right) + \omega^2 \rho u_z(\mathbf{r}) = 0. \tag{B 16}$$

As it can be seen, the obtained relation has the same form of equation (B 9), (B 13) and, in particular, it clearly emerges that equation (B 16) exactly matches the prototype problem in equation (B 3), provided that

$$(\tilde{\Gamma}(\mathbf{r}), p_1(\mathbf{r}), p_2(\mathbf{r})) \rightarrow \left( u_z(\mathbf{r}), \frac{1}{G(\mathbf{r})} \equiv \frac{1}{G}, \rho(\mathbf{r}) \equiv \rho \right). \tag{B 17}$$

Again, solving equation (B 16) for $u_z(\mathbf{r})$ suffices for the complete determination of the physical fields involved in the problem, i.e. displacement field and derived stress field within the medium. Similar considerations apply for the particular case of elastic waves propagation considered in the present paper, i.e. unidirectional time-harmonic compression waves. Again, the wave propagation equation exactly matches the general expression in (B 3) just by substituting

$$(\tilde{\Gamma}(\mathbf{r}), p_1(\mathbf{r}), p_2(\mathbf{r})) \rightarrow \left( u_j(x), \frac{1}{E_j}, \rho_j \right), \tag{B 18}$$

being the index $j = 1, 2$ corresponding to the two layers considered (cf. §3).

As a conclusion, it can be said that the aforementioned problems, i.e. transverse TE and TM electromagnetic waves, anti-plane shear waves and unidirectional compression waves, can be described by the same wave propagation equation (B 3) whose solution allows not only the complete determination of the physical fields involved, but also the possibility to make similar considerations for the different considered cases. Obviously, an appropriate interpretation of the quantities involved has to be made.

Having established this so-called elastic-electromagnetic equivalence [45,49,50] is of great importance because we can apply the modelling techniques developed for the solution of one problem to the others. In [51], for example, the homogenization technique, which is adhered to in this paper (cf. §2), has been applied for the analysis of periodic metamaterials made of magnetodielectric inclusions. Based on a Floquet-based approach, the authors derived closed-form expressions for the constitutive parameters, i.e. effective permittivity and permeability as a function of the inclusions geometry. It emerged the possibility to tailor the range of frequencies for which the effective properties become negative so that, as a consequence, the corresponding waves cannot propagate. In the context of electromagnetic waves propagation, the idea of analysing band gaps according to the sign of the effective parameters, has also been adopted in [52]. This method, however, because of the similarity of the mathematical description, has also been used in the context of elastic waves propagation. Vondrejc *et al*. [53], in particular, focuses on strongly heterogeneous elastic composites and investigated the dependence of the band gaps distribution on the sign of the effective mass tensor. Based on this, an optimization problem to maximize band gaps is also proposed. Cross-property relations linking the elastic and electromagnetic wave characteristics can also be made, leading to the development of multifunctional materials able to interact with both elastic and electromagnetic waves [54]. This has been done in [54], where the effective electromagnetic (dynamic dielectric constant, wave speed and attenuation coefficient) and elastodynamic (bulk modulus, shear modulus and corresponding wave speed and attenuation coefficient) properties of two-phase composite media have been derived in the long-wavelength regime and linked to each other by using Fourier- space construction techniques. Finally, based on the electro-mechanical coupling, Rohan *et al*. [55] proposed a smart composite structure whose mechanical response can be actively controlled by imposing an external electrical field.